# Complex Mapping between Neural Response Frequency and Linguistic Units in Natural Speech

Yuran Zhang[1], Jiajie Zou[1], Nai Ding[1,2*]

[1] Key Laboratory for Biomedical Engineering of Ministry of Education,

College of Biomedical Engineering and Instrument Sciences,

Zhejiang University, Hangzhou 310027, China

[2] The MOE Frontier Science Center for Brain Science & Brain-machine Integration,

Zhejiang University, Hangzhou 310012, China

*Corresponding author

E-mail: ding_nai@zju.edu.cn (ND)

# Abstract

When listening to connected speech, human brain can extract multiple levels of linguistic units, such as syllables, words, and sentences. It has been hypothesized that the time scale of cortical activity encoding each linguistic unit is commensurate with the time scale of that linguistic unit in speech. Evidence for the hypothesis originally comes from studies using the frequency-tagging paradigm that presents each linguistic unit at a constant rate, and more recently extends to studies on natural speech. For natural speech, it is sometimes assumed that neural encoding of different levels of linguistic units is captured by the neural response tracking speech envelope in different frequency bands (e.g., around 1 Hz for phrases, around 2 Hz for words, and around 4 Hz for syllables). Here, we analyze the coherence between speech envelope and idealized responses, each of which tracks a single level of linguistic unit. Four units, i.e., phones, syllables, words, and sentences, are separately considered. It is shown that the idealized phone-, syllable-, and word-tracking responses all correlate with the speech envelope both around 3-6 Hz and below ~1 Hz. Further analyses reveal that the 1-Hz correlation mainly originates from the pauses in connected speech. The results here suggest that a simple frequency-domain decomposition of envelope-tracking activity cannot separate the neural responses to different linguistic units in natural speech.

# Introduction

Speech is a highly complex signal and linguistic theories have defined a hierarchy of speech units with distinct properties, such as phonemes, syllables, morphemes, words, phrases, and sentences (Fromkin & Rodman, 1983). A critical question in psycholinguistics and neurolinguistics is to test which of the linguistically defined units are psychologically real, i.e., being represented in the brain (Fodor & Bever, 1965; Halle et al., 1978; Murphy, 2015). In other words, are the abstract units defined by linguists extracted by the human brain during natural speech comprehension? If these linguistic units are extracted, how are they represented by neural activity? One hypothesis is that the neural response to a linguistic unit should have the same time scale as the unit in the speech stimulus (Ding et al., 2016). According to this hypothesis, the neural response to a multisyllabic word should live longer than the response to a component syllable, and similarly, the response to a multi-word sentence should last longer than the response to each of the component words. Evidence for the hypothesis originally comes from frequency-tagging experiments in which each linguistic unit is presented at a constant frequency (Burroughs et al., 2021; Glushko et al., 2022; Jin et al., 2020; Kalenkovich et al., 2022; Lu et al., 2019). In these experiments, to isolate the neural response to a superordinate linguistic unit that is internally constructed, acoustic cues related to the unit, e.g., word or phrase, are removed using synthesized speech (see also, e.g., (Nozaradan et al., 2011) for investigation of musical rhythms using a similar approach).

The frequency-tagging paradigm is powerful to isolate the neural response to a linguistic unit, but the synthesized isochronous speech is clearly unnatural. Therefore, researchers have been actively developing new methods to extract the neural responses to different linguistic units when listening to natural speech (Brodbeck et al., 2018; Broderick et al., 2018; Keitel et al., 2018). One method is to directly generalize the frequency-domain analysis in the frequency tagging paradigm to study natural speech: In natural speech, different linguistic units are clearly not isochronous but they tend to be quasi-rhythmic. For example, the mean syllable rate of speech is around 4-5 Hz (Coupé et al., 2019; Ding et al., 2017; Greenberg et al., 2003; Poeppel & Assaneo, 2020; Y. Zhang et al., 2023). Since a phrase or a sentence typically has multiple syllables, its occurrence frequency is below the syllable rate and, according to corpus analyses, the rate of the intonational phrase is roughly around 1 Hz (Inbar et al., 2020; Stehwien & Meyer, 2021). Therefore, in a number of studies, the neural responses near 4-5 Hz and below 1 Hz are viewed as the syllabic-level and phrasal-level responses, respectively, and the word-level response has an intermediate frequency near 2 Hz (Coopmans et al., 2022; Kaufeld et al., 2020; Keitel et al., 2018; ten Oever et al., 2022). Furthermore, to ensure that the narrow-band neural responses are related to speech processing, these studies restrain the analysis to the neural response tracking the speech envelope. In the following, this frequency-domain approach is referred to as the multiscale-envelope tracking paradigm.

A critique of the multiscale-envelope tracking paradigm is that the actual duration of a phrase is highly variable (Kazanina & Tavano, 2022). In other words, the neural response tracking phrases may not be a narrowband signal. Nevertheless, it can also be argued that the response near 1 Hz captures a main component of the phrase response. Another issue is that the paradigm only analyzes the neural response tracking the speech envelope, a basic acoustic feature of speech. For example, the phrase-level neural response extracted in this paradigm is a 1-Hz neural response tracking the 1-Hz speech envelope, instead of linguistically defined phrases. In other words, the response analyzed in the multiscale-envelope tracking paradigm tracks low-level acoustic features that are considered as confounding factors and eliminated in the frequency-tagging paradigm. Therefore, the neural responses analyzed in these two paradigms are orthogonal to each other (Kaufeld et al., 2020). Both external sensory features and internal linguistic knowledge, however, are critical to speech comprehension and both deserve investigation.

Here, we ask whether the analysis techniques for the frequency-tagging paradigm can safely generalize to the multiscale-envelope tracking paradigm. We analyze whether the neural responses tracking different linguistic units, i.e., phones, syllables, words, and phrases, correlate with the speech envelope in separable narrow frequency bands, as is hypothesized by the multiscale-envelope tracking paradigm. Using a straightforward simulation, it is demonstrated that even if a neural response only tracks the onset of each phone, syllable, or word, it strongly correlates with the speech envelope below 1 Hz and the correlation can be even stronger than the correlation with the speech envelope around 4 Hz. Further analyses show that the very low-frequency, i.e., <~1 Hz, correlation between phone/syllable/word responses and speech envelope is caused by pauses in speech.

## Methods

**Speech materials**

The analysis was based on audiobooks from two speech corpora, i.e., WenetSpeech (B. Zhang et al., 2022) and GigaSpeech (Chen et al., 2021). WenetSpeech contained 41 hours of recordings of Chinese audiobooks, and GigaSpeech contained 85 hours of recordings of English audiobooks. More details about the selection of audiobook corpora could be found in Y. Zhang et al. (2023).

The duration of speech recordings ranged from 15 s to 691 s (M = 34 s, SD = 30 s). Additionally, we also analyzed synthesized speech used in Ding et al. 2016 for comparison. The analysis included 20 Chinese speech sequences and 30 English speech sequences. Each sequence contained 40 or 48 4-syllable sentences. For Chinese, syllables are presented at a constant rate of 4 Hz, and for English, syllables are presented at a constant rate of 3.125 Hz.

**Extraction of speech envelope and linguistic units**

The speech envelope was extracted by half-wave rectifying the speech waveform and low-pass filtering the rectified waveform (FIR filter, cut-off frequency 20 Hz). The boundaries between phones and syllables were extracted based on the transcriptions using the methods described in (Y. Zhang et al., 2023). The boundaries between words and sentences were available from the transcription for English. For Chinese, only the boundaries between sentences were available from the transcription and the word boundaries were obtained using a word segmentation tool (Che et al., 2021). The onset of a word/sentence was defined as the onset of its first phone.

An idealized neural response encoding just one level of linguistic unit was simulated by a 0-1 sequence whose value was 1 only at the onset of each target linguistic unit (Figure 1). The target linguistic unit could be a phone, a syllable, a word, or a sentence. Such an idealized neural response could be viewed as the output of a neuron that could ideally detect each target unit and generate a spike. The actual MEG/EEG responses were not spikes. However, a smoothed response could easily be obtained by convolving the idealized response with a smooth window, e.g., a 200-ms Hann window (see Figure 3A). Furthermore, the idealized response only encoded unit onset, and we also considered an accumulator model for which the response amplitude linearly ramped up within the duration of a unit (see Figure 3A).

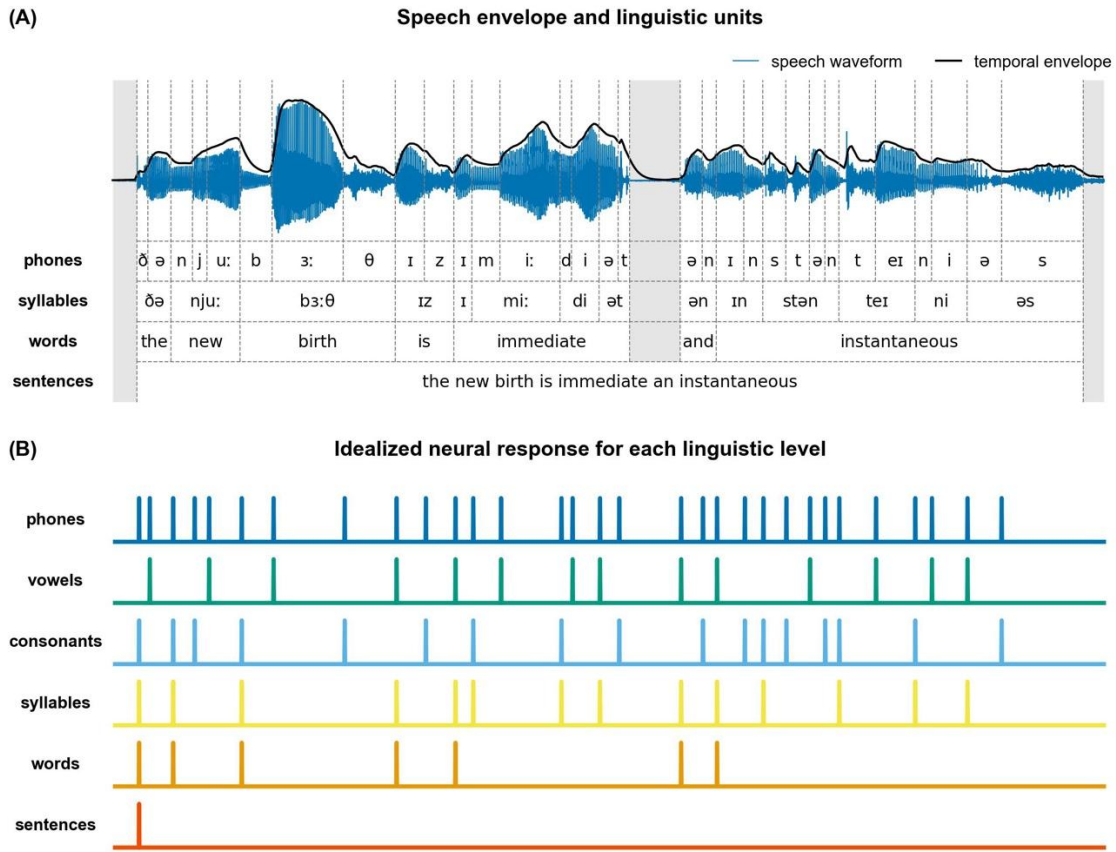

**Figure 1. Speech envelope and different levels of linguistic units.** (A) Speech envelope and linguistic units. The boundaries of phones, vowels, consonants, syllables, words, and sentences are marked by dotted lines and pauses are shaded. (B) Onset sequence of each level of linguistic unit, which is 1 at the onset of a unit and 0 otherwise.

Finally, pauses in the speech were defined as periods in which the amplitude was less than 10% of the maximum value and the duration was longer than 100 ms. In a control analysis, all silent periods were removed from the envelope and idealized neural response for a linguistic unit.

**Phase coherence spectrum**

The phase coherence spectrum was employed to reveal the frequency bands in which the speech envelope was correlated with idealized neural activity that only responds to a target level of linguistic unit. In this analysis, the speech envelope and idealized neural activity were both segmented into 10-s nonoverlapping epochs and transformed into the frequency domain using the DFT. The DFT coefficients of the syllable onset sequence and the broadband envelope were referred to as $\sigma(f)$ and $env(f)$, respectively. The coherence spectrum $P(f)$ was calculated using the following equation:

$$P(f) = \left(\frac{\sum_{i=1}^{N} \sin(\sigma_i(f) - env_i(f))}{N}\right)^2 + \left(\frac{\sum_{i=1}^{N} \cos(\sigma_i(f) - env_i(f))}{N}\right)^2$$

where $N$ was the number of epochs in each corpus. A chance-level coherence spectrum was calculated by randomly pairing the syllable onset sequence in one epoch and the speech envelope from another epoch. The chance-level coherence spectrum was always subtracted from the phase coherence spectrum.

## Results

We employed the phase coherence spectrum to analyze in which frequency bands the speech envelope correlated with idealized neural activity that only responds to the onset a target level of linguistic unit (Figure 1). The target unit could be a phone, a syllable, a word, or a sentence.

For naturally produced audiobooks, the phase coherence spectrum was shown in left panels of Figure 2A for both Chinese and English. For vowels, consonants, syllables, and words, the phase coherence spectrum revealed phase locking between speech envelope and idealized neural responses in two frequency bands. One frequency band was between 3 and 6 Hz, and the other band was below 1 Hz. For phones and sentences, the phase coherence spectrum showed a single peak below 1 Hz. Similar results were obtained using alternative ways to simulate the idealized response to each level of linguistic unit (Figure 3).

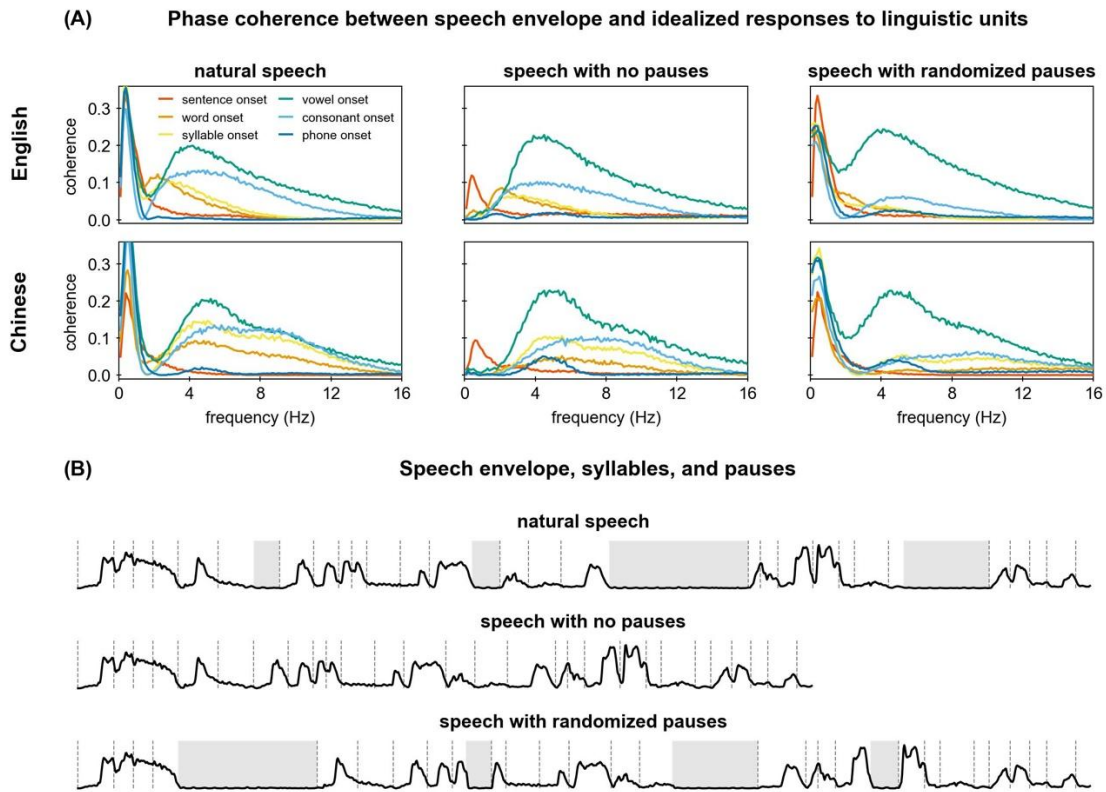

**Figure 2. Phase coherence spectrum for naturally produced audiobooks.** (A) Phase coherence between speech envelope and an idealized neural response for each linguistic unit. The left panels show the phase coherence spectrum for natural speech. For vowels, consonants, syllables, and words, high coherence is observed both between 3-6 Hz and below 1 Hz. For phones and sentences, a single peak is observed below 1 Hz. The middle panels show the phase coherence spectrum after pauses in speech are removed. A low-frequency peak below 1 Hz is only observed for sentences. The right panels show the phase coherence spectrum when the temporal locations of pauses are randomized. A low-frequency peak below 1 Hz is observed for all linguistic units. (B) Illustration of how pauses are removed or randomized for the syllable-tracking activity. The boundaries of syllables are marked by dotted lines and pauses are shaded. For the randomized condition, a pause is removed from its original position and inserted after a randomly chosen unit.

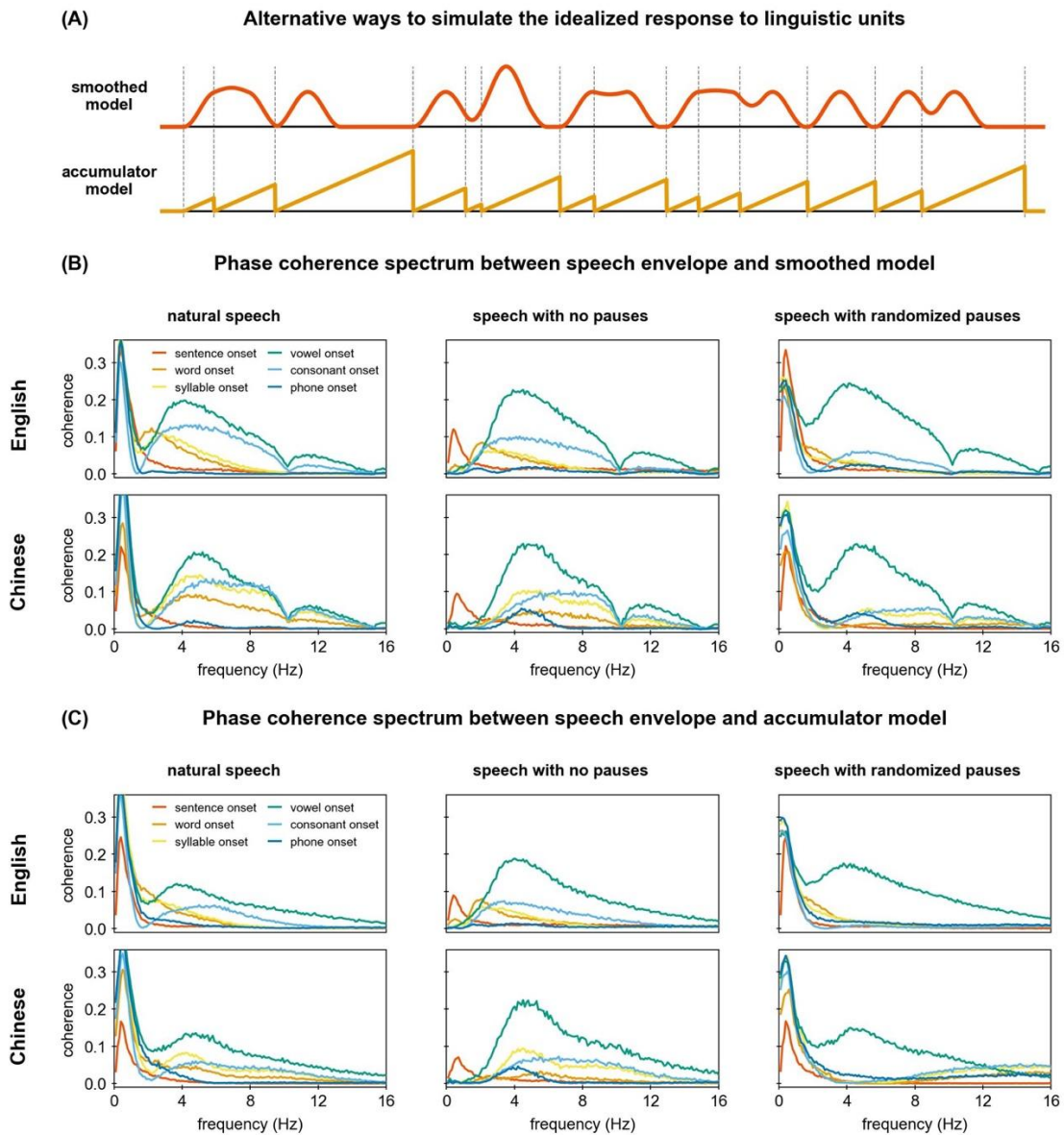

**Figure 3. Alternative ways to simulate the idealized response to linguistic units.** (A) Examples for the syllable-tracking response. The smoothed model is obtained by smoothing the syllable onset sequence using a 200-ms Hann window. For the accumulator model, the response linearly ramps up after the onset of a unit and resets to 0 at the onset of the next unit. The boundaries of units are marked by dotted lines. (BC) Phase coherence calculated based on the smoothed model and accumulator model. Same conventions as in Figure 2.

Phones and syllables were typically short units within 500 ms (Greenberg et al., 2003), what were the factors driving the very low-frequency peak below 1 Hz? One possibility was the pauses in speech, which could last for much longer in duration than phones and syllables. To test this hypothesis, we artificially removed all pauses in speech and found that the 0-Hz peak disappeared when the pauses were removed (middle panel of Figure 2A). Since the low-frequency peak in the phase coherence spectrum was indeed caused by pauses, next we investigated whether it was sensitive to the existence of pauses or the time intervals between pauses. In this analysis, we randomized the time intervals between pauses by taking out each pause and inserting it behind a randomly chosen linguistic unit. For speech with randomized pauses, the low-frequency peak remained (right panel of Figure 2A). Therefore, the existence, instead of the temporal position of pauses, generated the low-frequency phase coherence between speech envelope and the neural responses tracking linguistic units.

Finally, we also analyzed the isochronous synthesized speech used in frequency tagging studies (Ding et al., 2016). For the synthesized speech, the phase coherence spectrum only shows a peak at the syllable rate and harmonically related frequencies for all linguistic units (Figure 4).

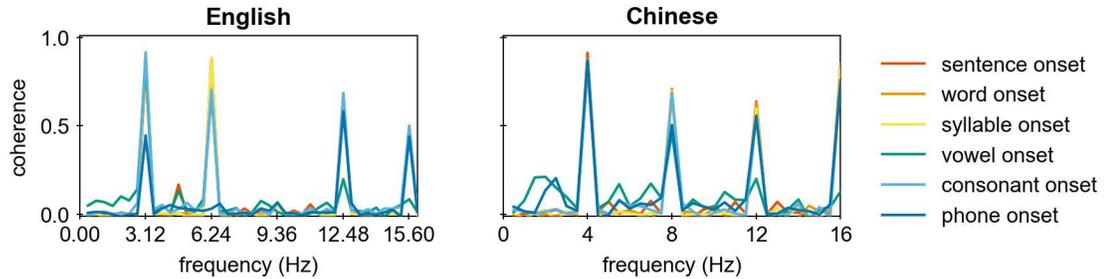

**Figure 4. Phase coherence spectrum for synthesized isochronous speech used in Ding et al., (2016).** For English and Chinese, syllables are presented at 3.12 Hz and 4 Hz, respectively. Strong phase coherence is only observed at the syllable rate and harmonically related frequencies.

## Discussion

We analyze how each level of linguistic unit is related to the speech envelope in the frequency domain using the phase coherence spectrum analysis, a method widely used to characterize the synchronization between speech envelope and neural responses (Cavanagh et al., 2009; Doelling et al., 2014; Peelle et al., 2013; Zou et al., 2021). The results here suggest that, for natural speech, each level of the linguistic unit often correlates with two frequency bands in the speech envelope. Only when the pauses in speech are removed, each level of linguistic unit correlates with a single frequency band in the speech envelope. Compared with larger linguistic units, smaller units tend to correlate a lower frequency region in the speech envelope, but there is heavily overlapping in frequency across different linguistic levels.

In the following, we discuss how the results contribute to our interpretation of the multiscale-envelope tracking paradigm. The multiscale-envelope tracking paradigm relies on a basic assumption that linguistic units of different sizes linearly map onto different regions along the spectrum of speech envelope (ten Oever et al., 2022). In other words, smaller/larger linguistic units separately map onto higher/lower-frequency components in the speech envelope. The current study, however, demonstrates that this assumption is not true for phones, syllables, and words (Figure 2). The current results show that neural activity that purely tracks individual phones, syllables, or words can appear to track very low-frequency (~1 Hz) speech envelope, violating the assumption in the multiscale-envelope tracking paradigm that neural tracking of very low-frequency envelope reflects phrasal-level processing.

The low-frequency component of the idealized syllable-tracking response can be avoided when the syllables are presented at a constant rate (Figure 4). In the frequency tagging paradigm (Ding et al., 2016), the original purpose to present syllables at a constant rate is to avoid the acoustic cues for word and phrase boundaries and the neural responses to these cues. The current simulation study, however, shows that the paradigm can also avoid that the power of a syllable-tracking response leaks to low-frequency regions and interferes with the measurement of phrasal-level responses.

Here, we simulate a neural response that precisely tracks and only tracks a single level of linguistic unit. The simulation is idealized and is not supposed to be biophysically plausible since we are not aware of any hypothesis about which biophysical mechanisms can extract multiple levels of linguistic units based on the speech input. Furthermore, the multi-scale envelope tracking paradigm, as well as the frequency-tagging paradigm (Ding et al., 2016), only proposes neural markers for the mental processing of linguistic units, but have not yet probed at the biophysical level how the neural markers can be generated.

The correlation between syllable-tracking activity and the low-frequency (~1 Hz) speech envelope is caused by the pauses in speech and largely disappears when all pauses are removed. It should be noted that this phenomenon is not related to any neural response that encodes the pauses in speech, since in the current simulation the neural activity only tracks the syllables and does not directly encode any other features, including the acoustic onsets induced by pauses or the speech envelope. In other words, the simulated response encodes each syllable in exactly the same way, whether the syllable is preceded/followed by a pause or not. Consequently, the syllable encoding mechanism is the same across the 3 columns in Figure 2, and the distinctions in the phase coherence solely arise from the distinctions in the input rhythm. The actual neural response is actually more complicated, since some components of the neural response can certainly track the low-frequency speech envelope and neural activity is also sensitive to pauses in speech (Ding & He, 2016; Keitel et al., 2018).

Syllable-tracking neural activity is a potential contributor to neural tracking of the very low-frequency speech envelope, but it is certainly not the only contributor. For example, it is well known that neural activity can track very slow changes in the speech envelope, even for non-speech sounds (Peter et al., 2022; Reetzke et al., 2021). Even for speech, EEG studies have shown that neural tracking of the 1-Hz envelope is enhanced when listening to an unknown language compared to the native language (Reetzke et al., 2021; Zou et al., 2019). Furthermore, neural activity can be sensitive to pauses in speech, and such neural activity can also contribute to the very low-frequency envelope-tracking response. For example, speech onsets can trigger an event-related response (Anurova et al., 2022) and a recent MEG study demonstrates that low-frequency (<1 Hz) neural tracking of speech envelope is strongest near speech onsets, i.e., after pauses (Chalas et al., 2023).

In summary, the current study demonstrates that the speech envelope correlates with the onsets of multiple levels of linguistic units, but the coherence between speech envelope and different linguistic units is not well separated in frequency. For natural speech, all linguistic levels correlate with very low-frequency speech envelope and therefore the neural response correlating with a very low-frequency speech envelope can potentially include at least 3 components: (1) auditory responses tracking the sound envelope, (2) responses to phrase-related acoustic cues; (3) responses tracking the onsets of vowels, consonants, syllables, and words, as is demonstrated by the current study. Therefore, it is not well justified to directly attribute low-frequency envelope-tracking activity as a neural correlate of phrase processing. Frequency tagging is a paradigm that can avoid the very low-frequency envelope-tracking response (Figure 4) and efficiently isolate neural processing of higher-level linguistic information. For natural speech, methods such as the temporal response function (TRF) do not assume specific spectral properties of the linguistic responses are potentially more promising to capture the neural encoding of linguistic features in natural speech (Aertsen & Johannesma, 1981; Brodbeck & Simon, 2020; Di Liberto et al., 2015; Theunissen et al., 2001).

## Data Availability Statement

The speech corpora used for analyses are publicly available and can be accessed at https://github.com/SpeechColab/GigaSpeech and https://wenet.org.cn/WenetSpeech. The code used to analyze the data can be accessed at https://github.com/austin-365/ms-tools.

## Author Contributions

Yuran Zhang: Data Curation; Formal Analysis; Investigation; Methodology; Resources; Validation; Visualization; Writing - Original Draft Preparation; Writing - Review & Editing. Jiajie Zou: Conceptualization; Methodology; Resources; Software; Supervision. Nai Ding: Conceptualization; Funding Acquisition; Investigation; Methodology; Project Administration; Supervision; Validation; Visualization; Writing - Original Draft Preparation; Writing - Review & Editing.

## Funding Information

This work was partly supported by the National Natural Science Foundation of China (no. 32222035) and the Key R&D Program of Zhejiang (no. 2022C03011).

Doelling, K. B., Arnal, L. H., Ghitza, O., & Poeppel, D. (2014). Acoustic landmarks drive delta-theta oscillations to enable speech comprehension by facilitating perceptual parsing. *NeuroImage*, *85*, 761–768. https://doi.org/10.1016/j.neuroimage.2013.06.035

Fodor, J. A., & Bever, T. G. (1965). The psychological reality of linguistic segments. *Journal of Verbal Learning and Verbal Behavior*, *4*(5), 414–420. https://doi.org/10.1016/S0022-5371(65)80081-0

Fromkin, V., & Rodman, R. (1983). *An introduction to language*. Linguistic Society of America. https://doi.org/10.2307/413657

Glushko, A., Poeppel, D., & Steinhauer, K. (2022). Overt and implicit prosody contribute to neurophysiological responses previously attributed to grammatical processing. *Scientific Reports*, *12*(1), 14759. https://doi.org/10.1038/s41598-022-18162-3

Greenberg, S., Carvey, H., Hitchcock, L., & Chang, S. (2003). Temporal properties of spontaneous speech – a syllable-centric perspective. *Journal of Phonetics*, *31*(3–4), 465–485. https://doi.org/10.1016/j.wocn.2003.09.005

Halle, M. E., Bresnan, J. E., & Miller, G. A. (1978). *Linguistic theory and psychological reality*. Massachusetts Inst of Technology Pr. https://doi.org/10.2307/414117

Inbar, M., Grossman, E., & Landau, A. N. (2020). Sequences of Intonation Units form a ~1 Hz rhythm. *Scientific Reports*, *10*(1), 15846. https://doi.org/10.1038/s41598-020-72739-4

Jin, P., Lu, Y., & Ding, N. (2020). Low-frequency neural activity reflects rule-based chunking during speech listening. *ELife*, *9*, e55613. https://doi.org/10.7554/eLife.55613

Kalenkovich, E., Shestakova, A., & Kazanina, N. (2022). Frequency tagging of syntactic structure or lexical properties; a registered MEG study. *Cortex*, *146*, 24–38. https://doi.org/10.1016/j.cortex.2021.09.01219